\documentclass{article}
\usepackage{ams}
\usepackage{graphicx}

\begin{document}
\title{Motion of pole-dipole and quadrupole particles in non-minimally coupled theories of gravity}
\author{Morteza Mohseni\thanks{E-mail address:m-mohseni@pnu.ac.ir}\\
\small Physics Department, Payame Noor University, 19395-4697
Tehran, Iran}
\date{}
\maketitle
\begin{abstract}
We study theories of gravity with non-minimal coupling between polarized media with pole-dipole and quadrupole moments and an arbitrary 
function of the space-time curvature scalar $R$ and the squares of the Ricci and Riemann curvature tensors. We obtain the general form of the 
equation of motion and show that an induced quadrupole moment emerges as a result of the curvature tensor dependence of the function coupled to 
the matter. We derive the explicit forms of the equations of motion in the particular cases of coupling to a function of the curvature scalar  
alone, coupling to an arbitrary function of the square of the Riemann curvature tensor, and coupling to an arbitrary function of the 
Gauss-Bonnet invariant. We show that in these cases the extra force resulting from the non-minimal coupling can be expressed in terms of the 
induced moments.  

\vspace{1cm}
PACS: 04.20.Cv,04.20.Fy,04.40.-b,04.50.-h
\end{abstract}
\section{Introduction}
Various extensions of the general theory of relativity have been proposed in recent years to explain astrophysical observations like the accelerated expansion of the Universe and rotation curves of spiral galaxies. A widely known class of such proposals comprises variants of the so-called $f(R)$ theories of gravity; see \cite{capozziello,sotifara,tue} for reviews. Even these theories have been themselves subject to 
different modifications in the hope of obtaining a more satisfactory picture, namely, by considering non-minimal couplings with matter 
\cite{goner,odintsov,nojiri,corda}. In the models investigated in these references some powers of the curvature scalar $R$ have been 
coupled to the matter Lagrangian which have resulted in modified equations of motion. A more general case was considered in \cite{odin2} in 
which an arbitrary function of the curvature scalar non-minimally coupled with the matter Lagrangian. Taking perfect fluids and scalar fields 
as matter fields, several implications of such a coupling were subsequently studied in 
\cite{valer,paramos,bertolami,sotiriou,harko,laia,sequeira,nesseris,coco}. The effects of such a coupling on the dynamics of test 
particles have been studied in \cite{bohmer} where it has been shown that an extra force emerges from the non-minimal coupling. The explicit 
form of the force depends on the form of the  matter Lagrangian and it has been shown in  \cite{lobo,faraoni} that for the case of the perfect 
fluid for which there are several equivalent Lagrangians, one may end up with vanishing or non-vanishing extra forces depending on the chosen 
form of the Lagrangian. In \cite{faro}, it has been argued that the Lagrangians which are equivalent for a non-coupled perfect fluid are not in 
fact equivalent when the perfect fluid couples non-minimally to curvature. The author of \cite{harco} has argued that the form of the coupling 
fixes the matter Lagrangian.

The extra force affects the world-lines of test particles. In fact within the framework of models with non-minimal coupling, test particles move 
along non-geodesic trajectories in general \cite{bohmer}. Taking the effect of extra force into account, the motion of test pole-dipole 
particles, i.e. test bodies with certain microstructures, has been studied in \cite{puetzfeld} by using the method of multipoles 
\cite{obukhov07,puetzfeld08}. In this work we consider the motion of pole-dipole and quadrupole particles in extended gravity models with more 
general coupling to matter. To this end we consider a class of matter fields with a general dependence on a tetrad field, its covariant 
derivative, a velocity field, curvature tensors, and a generic set of matter fields, coupled to $f(R)$ gravity action. In particular we show 
that when the Lagrangian depends on curvature tensor through non-minimal coupling to geometry, induced quadrupole moments emerges. We obtain the 
equations of motion for media with pole-dipole and quadrupole moments non-minimally coupled to an arbitrary function of the curvature scalar and 
as a special case we consider a Weyssenhoff-type spinning fluid from which we extract the equations of motion of pole-dipole particles. The 
method can then be easily generalized to particles with electric charges. We also investigate the motion of test bodies non-minimally coupled to 
an arbitrary function of the square of the Riemann curvature tensor or to an arbitrary function of the Gauss-Bonnet invariant and show that the 
extra forces can be expressed in terms of the induced quadrupole moments. The problem of induced quadrupole moments in media due to an applied 
gravitational field has been studied in \cite{sze,mon,suen1,suen2,hu} in the context of general relativity. The study of motion of quadrupole 
particles is well motivated by their role in gravitational radiation. Several examples of motion of quadrupole particles in curved space-times 
of general relativity may be found in \cite{gr,bini1,bini2,bini3,bini4}. The effect of multipole moments of planets on bending of light rays has 
been discussed in \cite{kopi6}, and it has been suggested in \cite{crosta,kopi7} that such effects may be detected by using microarcsecond 
astronomical interferometers.  

In the subsequent sections we first give a brief review of the polarized media as formulated in \cite{israel,bailey} and then deploy it to 
obtain the equations of motion of polarized media with pole-dipole and quadrupole moments non-minimally coupled to an arbitrary function  
constructed out of the curvature scalar and the squares of the Ricci and Riemann curvature tensors. We subsequently consider three particular 
cases, the case of non-minimal $f(R)$ gravity, the case of coupling with an arbitrary function of the Riemann curvature tensor squared, and the 
case of coupling to an arbitrary function of the Gauss-Bonnet invariant. In the first case, our investigation results in equations of motion of 
particles with pole-dipole and quadrupole moments in non-minimal $f(R)$ gravity. In the second and third cases in which for simplicity no 
intrinsic moments (i.e. those originating from the structure of the body) are assumed, we show that the extra force of the non-minimal coupling 
can be expressed in terms of the induced moments. In the last section we discuss the results. Throughout the work we use the following 
(anti)symmetrization conventions: $A^{\cdots(\mu\nu)\cdots}\equiv A^{\cdots\mu\nu\cdots}+A^{\cdots\nu\mu\cdots}$, 
$A^{\cdots(\mu\cdots}B^{\cdots\nu)\cdots}\equiv A^{\cdots\mu\cdots}B^{\cdots\nu\cdots}+A^{\cdots\nu\cdots}B^{\cdots\mu\cdots}$, and 
$A^{\cdots[\mu\cdots}B^{\cdots\nu]\cdots}\equiv A^{\cdots\mu\cdots}B^{\cdots\nu\cdots}-A^{\cdots\nu\cdots}B^{\cdots\mu\cdots}$.       

\section{Particles with pole-dipole and quadrupole moments in General Relativity}
It has been shown in \cite{israel,bailey} that by starting from a Lagrangian with the general form ${\mathcal L}(u^\mu,e^\mu_a,{\dot 
e}^\mu_a,{R^\mu}_{\alpha\beta\gamma})$ in which $u^\mu$ is a velocity field, $e^\mu_a$ is a tetrad field satisfying $e^a_\mu 
e^b_\nu\eta_{ab}=g_{\mu\nu}$, with $\eta_{ab}=\mbox{diag}(-1,1,1,1)$, ${\dot e}^\mu_a=u^\alpha\nabla_\alpha e^\mu_a$, and 
${R^\mu}_{\alpha\beta\gamma}=\partial_\alpha\Gamma^\mu_{\beta\nu}-\partial_\beta\Gamma^\mu_{\alpha\nu}
+\Gamma^\mu_{\alpha\delta}\Gamma^\delta_{\beta\nu}-\Gamma^\mu_{\beta\delta}\Gamma^\delta_{\alpha\nu}$ is the Riemann curvature tensor, one 
can arrive at the following energy-momentum tensor
\begin{equation}\label{q4}
T_{\mu\nu}=p_\mu u_\nu+{\mathcal P}h_{\mu\nu}-B_{\mu\nu}+A_{\mu\nu}
\end{equation}
describing a polarized fluid with pole-dipole and quadrupole moments. Here $u^\mu$ is the four-velocity of a fluid element, $p^\mu$ its 
four-momentum, ${\mathcal P}$ is the pressure, 
\begin{eqnarray}
h_{\mu\nu}&=&g_{\mu\nu}+u_\mu u_\nu,\label{kl1}\\
B_{\mu\nu}&=&\frac{1}{2}\nabla_\alpha({s^\alpha}_{(\mu}u_{\nu)}+s_{\mu\nu}u^\alpha),\label{kl2}\\
A_{\mu\nu}&=&q_{\mu\alpha\beta\gamma}{R_\nu}^{\alpha\beta\gamma}-3q_{\nu\alpha\beta\gamma}{R_\mu}^{\alpha\beta\gamma}
-2\nabla_\beta\nabla_\alpha q^\alpha{}_{(\mu\nu)}{}^\beta\label{kl2a}
\end{eqnarray}
where $s^{\mu\nu}$ is the fluid spin density and 
\begin{equation}\label{p1}
{q^\mu}_{\alpha\beta\gamma}=\frac{\partial{\mathcal L}}{\partial{{R_\mu}^{\alpha\beta\gamma}}}
\end{equation}
is the quadrupole moment tensor, which is assumed to share the symmetries of the Riemann curvature tensor.  

The above energy-momentum tensor is symmetric provided the spin evolution equation
\begin{equation}\label{q5}
\nabla_\alpha(u^\alpha s^{\mu\nu})=p^\mu u^\nu-p^\nu u^\mu-4R^{[\mu\alpha\beta\gamma}{q^{\nu]}}_{\alpha\beta\gamma}
\end{equation}
holds. The translational equation of motion 
\begin{eqnarray}\label{q6}
&&u^\nu\nabla_\nu p^\mu+p^\mu\nabla_\nu u^\nu+h^{\mu\nu}\nabla_\nu{\mathcal P}+{\mathcal P}\nabla_\nu(u^\mu 
u^\nu)=-\frac{1}{2}{R^\mu}_{\nu\alpha\beta}u^\nu s^{\alpha\beta}+q_{\alpha\beta\gamma\delta}\nabla^\mu R^{\alpha\beta\gamma\delta}
\end{eqnarray}
can be obtained from $\nabla_\nu T^{\mu\nu}=0$ or by variation of world-lines \cite{bailey}. The so-called Frenkel condition
\begin{equation}\label{kl3}
u_\mu s^{\mu\nu}=0
\end{equation}
ensures that spin remains space-like. Imposing the energy conditions may put further restrictions on the spin tensor. By neglecting the 
quadrupole moments, the equations of motion of a spinning fluid (see e.g. \cite{mohseni1} and references therein) are recovered. By turning the 
spin off and setting the quadrupole moments equal to zero, the standard perfect fluid energy-momentum tensor is recovered from Eq.~(\ref{q4}).

As has been shown in \cite{israel,bailey}, the above procedure can be extended to Lagrangians of more general dependence of the form 
\begin{equation}\label{pj}
{\mathcal L}\equiv{\mathcal L}(u^\mu,e^\mu_a,{\dot e}^\mu_a,\phi^A)
\end{equation}
in which
\begin{equation}\label{is1}
\phi^A=\{{R^\mu}_{\alpha\beta\gamma},\nabla_{\mu_1}\cdots\nabla_{\mu_m}{R^\mu}_{\alpha\beta\gamma},\varphi^A,\nabla_\alpha\varphi^A\}
\end{equation}
with $\varphi^A$ ($A$ is a collective index) being a generic set of matter fields whose free field Lagrangians are also included in the total 
Lagrangian. In particular by taking $\varphi^A$ to be the electromagnetic field, the equations of motion for charged particles would be 
obtained. Thus in the special case where particles have only pole-dipole moments, the Mathisson-Papapetrou-Dixon \cite{dixon} and 
Dixon-Souriau equations \cite{sau} for charge-less and charged particles are recovered respectively. Multipole moments higher than the 
quadrupole moment are incorporated via dependence on covariant derivatives of the Riemann curvature tensor.   

\section{Multipole particles in non-minimal extended gravity}
The action for $f(R)$ gravity with non-minimal coupling to matter is given by \cite{bohmer}
\begin{equation}\label{q1}
S=\int\left(\frac{1}{2}f_1(R)+\{1+\lambda f_2(R)\}{\mathcal L}_m\right)\sqrt{-g}d^4x
\end{equation}
where ${\mathcal L}_m$ represents the matter Lagrangian, $\lambda$ is a coupling constant, and $f_1(R)$ and $f_2(R)$ are arbitrary functions of 
the curvature scalar $R=g^{\alpha\gamma}g^{\beta\delta}R_{\alpha\beta\gamma\delta}$. For $\lambda=0$ this reduces to the usual $f(R)$ gravity. 
As has been shown in \cite{bohmer}, one can obtain the following relation from the equations of motion associated with the above action
\begin{equation}\label{q2}
\nabla_\nu T^{\mu\nu}=\frac{\lambda f_2^\prime(R)}{1+\lambda f_2(R)}\left(g^{\mu\nu}{\mathcal L}_m-T^{\mu\nu}\right)\nabla_\nu R
\end{equation}
where $f^\prime_2(R)=\frac{df_2(R)}{dR}$, and $T^{\mu\nu}$ is the matter energy-momentum tensor
\begin{equation}\label{q3}
T^{\mu\nu}=\frac{2}{\sqrt{-g}}\frac{\delta({\mathcal L}_m\sqrt{-g})}{\delta g_{\mu\nu}}.
\end{equation}
Here we consider a more general extension of the above action with the following form
\begin{eqnarray}\label{eq1}
S=\int{\sqrt{-g}}\left\{\frac{1}{2}R+\frac{1}{2}f+(1+\lambda F){\mathcal L}\right\}d^4x
\end{eqnarray}    
where $f\equiv f(R,P,Q)$, $F\equiv F(R,P,Q)$, and $P=R_{\mu\nu}R^{\mu\nu}$, $R^{\mu\nu}$ being the Ricci tensor, $Q=R_{\mu\nu\alpha\beta} 
R^{\mu\nu\alpha\beta}$ is the Kretschmann invariant, and ${\mathcal L}$ is a matter Lagrangian (see \cite{trod} for a cosmological model with 
this type of action but with $\lambda=0$). In fact one may think of $(1+\lambda F){\mathcal L}$ as a Lagrangian of the type described by 
Eq.~(\ref{pj}). Various models including minimal or non-minimal $f(R)$ gravity, modified Gauss-Bonnet gravity, the so-called $f(R,{\mathcal G})$ 
models \cite{aleman,def} where ${\mathcal G}=R^2-4P+Q$ is the Gauss-Bonnet invariant, and the model introduced in \cite{mohseni2} can be 
recovered as special cases by choosing appropriate forms for $f$ and $F$. Functions with more general dependence on curvature invariants (e.g. 
those introduced in \cite{zerb}) may also be considered as toy models.

The equations of motion can be obtained by varying the above action with respect to metric components. Defining 
\begin{eqnarray*}
H^{\mu\nu}&=&(R^{\mu\nu}-\nabla^\mu\nabla^\nu+g^{\mu\nu}\square)f_R+2R^{\mu\alpha}R^\nu_\alpha f_P 
-\nabla_\alpha\nabla^{(\nu}(f_PR^{\alpha\mu)})+g^{\mu\nu}\nabla_\alpha\nabla_\beta(f_P R^{\alpha\beta})\\&&
+\square(f_PR^{\mu\nu})+2R^{\mu\alpha\beta\gamma}{R^\nu}_{\alpha\beta\gamma}f_Q
-2\nabla_\alpha\nabla_\beta(f_QR^{\alpha(\mu\nu)\beta})
\end{eqnarray*}
we have
\begin{equation}\label{k1}
G^{\mu\nu}+H^{\mu\nu}-\frac{1}{2}fg^{\mu\nu}=(1+\lambda F)T^{\mu\nu}-2\lambda{\mathcal H}^{\mu\nu}
\end{equation}
where $\square=\nabla_\mu\nabla^\mu$, $\frac{df}{dR},\frac{df}{dP},\frac{df}{dQ},\frac{dF}{dR},\frac{dF}{dP}$, and $\frac{dF}{dQ}$ stand for  
$f_R,f_P,f_Q,F_R,F_P$, and $F_Q$ respectively, and ${\mathcal H}^{\mu\nu}$ is obtained from $H^{\mu\nu}$ by substituting 
$$f_R\rightarrow{\mathcal L} F_R,\,\,f_P\rightarrow{\mathcal L}F_P,\,\,f_Q\rightarrow{\mathcal L}F_Q.$$ The matter equation of motion may be 
obtained by taking the divergence of both sides of Eq.~(\ref{k1}), which results in 
\begin{equation}\label{q91}
\nabla_\nu T^{\mu\nu}=\frac{-\lambda}{1+\lambda F}\left(\nabla_\nu F T^{\mu\nu}-2\nabla_\nu{\mathcal H}^{\mu\nu}\right).
\end{equation}
Now if one chooses a Lagrangian of the form given by Eq.~(\ref{pj}), the contributions from intrinsic pole-dipole and quadrupole moments can 
be tracked through $T^{\mu\nu}$ terms in the above equation. However even if we choose a Lagrangian involving no pole-dipole or quadrupole 
moments, quadrupole moment contributions would still appear, through $F_R,F_P$, and $F_Q$ terms, in the dynamics. Thus the non-minimal coupling 
induces quadrupole moments in the medium. The appearance of these induced moments is an essential feature of the models with non-minimal 
coupling between geometry and matter. It is a novel phenomenon not observed in other modified theories of gravity. To proceed further, we 
consider some rather interesting special cases of this model.

\section{Pole-dipole and quadrupole particles in non-minimal $f(R)$ gravity}
In this section we restrict ourselves to a particular case of the model described above where $f=f(R)$, $F=F(R)$, and ${\mathcal L}={\mathcal L}
(u^\mu,e^\mu_a,{\dot e}^\mu_a,{R^\mu}_{\alpha\beta\gamma})$; i.e. we consider particles with pole-dipole and quadrupole moments in non-minimal 
$f(R)$ gravity. 

Inserting Eq.~(\ref{q4}) into Eq.~(\ref{q91}) we obtain
\begin{eqnarray}\label{q7}
u^\nu\nabla_\nu p^\mu+p^\mu\nabla_\nu u^\nu+h^{\mu\nu}\nabla_\nu{\mathcal P}+{\mathcal P}\nabla_\nu(u^\mu u^\nu)+\frac{1}{2}
{R^\mu}_{\nu\alpha\beta}u^\nu s^{\alpha\beta}-q_{\alpha\beta\gamma\delta}\nabla^\mu R^{\alpha\beta\gamma\delta}=f^\mu
\end{eqnarray}
with
\begin{equation}\label{q8}
f^\mu=\frac{\lambda F_R\nabla^\nu R}{1+\lambda F}(\delta^\mu_\nu{\mathcal L}_m-p^\mu u_\nu-{\mathcal P}h^\mu_\nu+{B^\mu}_\nu-{A^\mu}_\nu)
\end{equation}
being the extra force resulting from the non-minimal coupling. There are some quadrupole contributions to this extra force coming from 
dependence on curvature. These induced moment contributions can be computed with the aid of the following relation
\begin{equation}\label{ht}
q^{\alpha\beta\gamma\delta}=(1+\lambda F)\frac{\partial\mathcal L}{\partial R_{\alpha\beta\gamma\delta}}+\frac{1}
{2}g^{[\alpha\gamma}g^{\beta]\delta}\lambda F_R{\mathcal L} 
\end{equation}
which is obtained from Eq.~(\ref{p1}) by replacing ${\mathcal L}$ with $(1+\lambda F){\mathcal L}$. In this equation the first term in the 
right-hand side is proportional to the medium intrinsic quadrupole moment while the second one represents the quadrupole moment induced as a 
result of the non-minimal coupling. One may use the above type of equations to obtain a rough estimate of the magnitude of these induced 
moments. In the case of the model considered in this section the magnitude of the induced quadrupole moment relative to the intrinsic one is of 
the same order as of $\displaystyle\frac{\lambda F_R}{1+\lambda F}$.
  
To obtain the equations of motion of particles with pole-dipole and quadrupole moments, in Eq.~(\ref{q7}) we set $${\mathcal P}=0,\,\, 
p^\mu=nP^\mu,\,\,s^{\mu\nu}=nS^{\mu\nu},\,\,{q^\mu}_{\alpha\beta\gamma}=n{Q^\mu}_{\alpha\beta\gamma}$$ and use $\nabla_\alpha(n u^\alpha)=0$, 
where $n$ is the particle number density and $P^{\mu}, S^{\mu\nu}$, and ${Q^\mu}_{\alpha\beta\gamma}$ are the single-particle momentum, spin, 
and quadrupole moment respectively. This results in
\begin{equation}\label{q7b}
u^\nu\nabla_\nu P^\mu=f^\mu-\frac{1}{2}{R^\mu}_{\nu\alpha\beta}u^\nu S^{\alpha\beta}+Q_{\alpha\beta\gamma\delta}\nabla^\mu 
R^{\alpha\beta\gamma\delta}.
\end{equation} 
If we neglect the quadrupole moments, this equation is the non-minimal $f(R)$ gravity counterpart of the Mathisson-Papapetrou-Dixon equation 
and agrees with the results of \cite{puetzfeld}. By turning the spin off, the non-minimal coupling of $f(R)$-perfect fluid considered in 
\cite{bohmer} is achieved. For pole-dipole particles with electric charges one can add the Maxwell energy-momentum tensor to the right-hand side 
of Eq.~(\ref{q4}) and proceed as above. This would result in the non-minimal $f(R)$ gravity counterpart of the Dixon-Souriau equations. 

A particular case of the model described above is the case where $F(R)=R$. If we also assume, for simplicity, that ${\mathcal P}=0$ and 
the medium has no intrinsic pole-dipole or quadrupole moments, Eq.~(\ref{q7}) together with Eqs.~(\ref{q8}) and (\ref{ht}) give 
\begin{eqnarray}\label{y6}
\nabla_\nu(u^\nu p^\mu)&=&\frac{\lambda\nabla^\nu R}{1+\lambda R}\left(\frac{1}{6}\delta^\mu_\nu{q^{\alpha\beta}}_{\alpha\beta}-p^\mu 
u_\nu\right)
\end{eqnarray}
in which the following term 
\begin{eqnarray*}
\frac{1}{6}{q^{\alpha\beta}}_{\alpha\beta}\nabla^\mu\ln(1+\lambda R)\approx\frac{\lambda}{6}{q^{\alpha\beta}}_{\alpha\beta}\nabla^\mu R.
\end{eqnarray*}
gives the contribution from the induced moment. One may omit the term $p^\mu u_\nu$ in the right-hand side of Eq.~(\ref{y6}) by projecting 
normal to $u^\mu$. Note that according to \cite{sotiriou} this particular toy model is not cosmologically viable.   
 
We further remark that, as has been shown in \cite{bailey}, in deriving the energy-momentum tensors of the type given in Eq.~(\ref{q4}), it is not necessary to have the explicit form of the Lagrangian. Thus it is basically possible to start with different Lagrangians and end up with the 
same energy-momentum tensor. In this sense, the case of spinning fluids is even more degenerate than the perfect fluid. On the other hand for 
the explicit form of the Lagrangians proposed in the literature, e.g in \cite{obukhov}, the extra force does not vanish. 

\section{Particles in non-minimal $f(Q)$ gravity}
Now we take $f=f(R)$, $F=F(Q)$, and for simplicity take ${\mathcal L}$ the same as that of a perfect dust; i.e. we consider particles with no 
intrinsic pole-dipole or quadrupole moments. Inserting these data back into Eq.~(\ref{q91}) and projecting the resulting equation normal to 
$u^\mu$, we get 
\begin{eqnarray}\label{q7a}
\rho u^\nu\nabla_\nu u^\mu=\frac{4\lambda}{1+\lambda F}h^\mu_\kappa\nabla_\nu\{R^{\kappa\alpha\beta\gamma}{R^\nu}_{\alpha\beta\gamma}
F_Q{\mathcal L}-\nabla_\alpha\nabla_\beta(F_Q{\mathcal L}(R^{\alpha(\kappa\nu)\beta})\}.
\end{eqnarray}
Projection parallel to $u^\mu$ would result in a continuity equation. In the above equation, one can express $F_Q$ in terms of the quadrupole 
moments ${q^\mu}_{\alpha\beta\gamma}$. In fact, by using Eq.~(\ref{p1}) with ${\mathcal L}$ being replaced by $(1+\lambda F(Q)){\mathcal L}$, we 
have 
\begin{equation}\label{pp1}
{q^\mu}_{\alpha\beta\gamma}=2\lambda{\mathcal L}F_Q{R^\mu}_{\alpha\beta\gamma}
\end{equation}
from which we obtain
\begin{equation}\label{h1}
\lambda{\mathcal L}F_Q=\frac{1}{2Q}q_{\mu\alpha\beta\gamma}R^{\mu\alpha\beta\gamma}.
\end{equation}
Inserting this back into Eq.~(\ref{q7a}) we will arrive at an equation which shows that the particles do not follow geodesics of the space-time 
due an extra force resulting from the non-minimal coupling. The extra force, which is proportional to the right-hand side of Eq.~(\ref{q7a}), is 
normal to the particle trajectories and involves quadrupole moments. Thus, as in the previous case, the non-minimal coupling induces quadrupole 
moments in the medium. The magnitude of these induced moments depend on the strength of the coupling (controlled by $\lambda$) and the function 
$F$. 

For the particular case of $F(Q)=Q$ we have
\begin{eqnarray}\label{q7ab}
\rho u^\nu\nabla_\nu u^\mu&=&\frac{2\lambda}{1+\lambda Q}h^\mu_\kappa\nabla_\nu\left\{R^{\kappa\alpha\beta\gamma}
{R^\nu}_{\alpha\beta\gamma}\frac{q_{\rho\sigma\tau\omega}R^{\rho\sigma\tau\omega}}{Q} 
-\nabla_\alpha\nabla_\beta\left(\frac{q_{\rho\sigma\tau\omega}R^{\rho\sigma\tau\omega}}
{Q}R^{\alpha(\kappa\nu)\beta}\right)\right\}
\end{eqnarray}
which shows explicitly dependence on the induced moments. 

\section{Particles in non-minimal Gauss-Bonnet gravity}
Another interesting model sharing the above mentioned property is the one in which a function of the Gauss-Bonnet invariant is coupled to 
matter. Such models are of special interest partly due to the fact that $F(R,{\mathcal G})$ are the only 
subclass of $F(R,P,Q)$ models which are free of ghosts \cite{comelli,navaro}.

By taking $f=f(R)$, $F=F({\mathcal G})$ and performing calculations similar to those in the previous section, we obtain 
\begin{eqnarray}\label{q7f}
\rho u^\nu\nabla_\nu u^\mu&=&\frac{4\lambda}{1+\lambda F}h^\mu_\kappa\nabla_\nu{\mathcal K}^{\kappa\nu},
\end{eqnarray} 
where 
\begin{eqnarray*}
{\mathcal K}^{\kappa\nu}&=&(R^{\kappa\nu}-\nabla^\kappa\nabla^\nu+g^{\kappa\nu}\square){\mathcal 
L}RF^\prime-4R^{\kappa\alpha}R^\nu_\alpha{\mathcal L}  
F^\prime+2\nabla_\alpha\nabla^{(\nu}({\mathcal L}F^\prime R^{\alpha\kappa)})-2g^{\kappa\nu}\nabla_\alpha\nabla_\beta({\mathcal L}F^\prime 
R^{\alpha\beta})\\&&-2\square({\mathcal L}F^\prime R^{\kappa\nu})+R^{\kappa\alpha\beta\gamma}{R^\nu}_{\alpha\beta\gamma}{\mathcal 
L}F^\prime-\nabla_\alpha\nabla_\beta({\mathcal L}F^\prime R^{\alpha(\kappa\nu)\beta})
\end{eqnarray*}
and $F^\prime$ stands for $\frac{dF(\mathcal G)}{d\mathcal G}$. Also from Eq.~(\ref{p1}) with $\mathcal L$ replaced by $(1+\lambda F)\mathcal 
L$ we have
\begin{equation}\label{h1a}
\lambda{\mathcal L}F^\prime=\frac{1}{2\mathcal G}q_{\mu\alpha\beta\gamma}R^{\mu\alpha\beta\gamma}
\end{equation}
which can be inserted back into the expression for ${\mathcal K}^{\kappa\nu}$ to express the right-hand side of Eq.~(\ref{q7f}) in terms of the 
induced quadrupole moments.
\section{Conclusions}
We have obtained equations of motion of polarized media with non-minimal coupling to an arbitrary function of the curvature scalar and the 
squares of the Ricci and Riemann curvature tensors. As a special case, we considered the non-minimal $f(R)$ coupling and obtained the equations 
of motion of particles with pole-dipole and quadrupole moments which agree with the results of \cite{puetzfeld} when the quadrupole 
moments are absent. Such equations might be used to study the motion of astrophysical objects with pole-dipole or quadrupole moments in the 
context of modified gravity with non-minimal coupling. 

We have shown that, for coupling to a function of rather general dependence as described above, the non-minimal coupling induces quadrupole 
moments. The magnitude of these induced moments depend on the strength of the coupling and the function coupled to the matter field. The extra 
force resulting from the coupling can be expressed in terms of the induced moments. We have shown this explicitly for the example case where the 
coupling function depends only on the Kretschmann invariant and also for the case where the matter Lagrangian couples to an arbitrary function 
of the Gauss-Bonnet invariant. By taking dependence on covariant derivatives of the Riemann curvature tensor, higher order intrinsic and induced 
moments can also be accounted for. 

The emergence of induced moments is an essential property differentiating non-minimal modified gravity theories from other modified gravity 
theories. These induced moments are model-dependent, and possible detection of them via solar system experiments such as the bending of light 
rays using high precession measurements like those provided by microarcsecond astronomical interferometers might be a test for these models. 
Keeping the interrelations between gravity and geometry in mind, the quadrupole moments induced by non-minimal coupling to geometry may be 
compared with the induced quadrupole moments from gravitational fields applied to the media. 
\section*{Acknowledgements}
I thank Donato Bini for helpful comments on an early draft of this manuscript. I am also indebted to two anonymous referees of the 
Physical Review D for valuable comments.

\end{document}